\documentclass{aa}
\usepackage{times}
 
\newcommand{\rxj}   {RX~J0146.9+6121~}

\newcommand{\ltsima} {$\; \buildrel < \over \sim \;$}
\newcommand{\simlt}  {\lower.5ex\hbox{\ltsima}}            
\newcommand{\gtsima} {$\; \buildrel > \over \sim \;$}
\newcommand{\simgt}  {\lower.5ex\hbox{\gtsima}}            
  
\newcommand{\be} {\begin{equation}}

\newcommand{\ee} {\end{equation}}
\newcommand{\etal}{{\it et al. }}

\newcommand{\BSAX}{{\em Beppo}SAX} 
\newcommand{\RXTE}{{\em Rossi}XTE}
\newcommand{\bc}{\begin{center}}
\newcommand{\ec}{\end{center}}
\newcommand {\rchisq}{$\chi_{\nu} ^{2}$}

\def \hcm {\hbox {\ifmmode $ atoms cm$^{-2}\else atoms cm$^{-2}$\fi}}
\def\deg {^\circ}

\begin{document}
 
\thesaurus{06(08.09.2 \object{RX~J0146.9+6121 };  ;  )}
 
\title
{\RXTE\ and \BSAX\ observations of   
the Be/neutron star system \rxj }
 
\author{ S.~Mereghetti\inst{1}, A.~Tiengo\inst{1,2}, 
G.L.~Israel\inst{3, }\thanks{Affiliated to I.C.R.A.} and  L.~Stella\inst{3}$^{, \star}$
}
 
\institute{
{Istituto di Fisica Cosmica ``G.Occhialini'',
via Bassini 15, I-20133 Milano, Italy}  \and
{Dipartimento di Fisica, Universit\`a di Milano, Via Celoria 16, I-20133
Milano, Italy} \and
{Osservatorio Astronomico di Roma, Via dell'Osservatorio 2,
I-00040 Monteporzio Catone (Roma), Italy}
}
 
\offprints{S.Mereghetti, sandro@ifctr.mi.cnr.it}
 
\date{Received July 1999/ Accepted December 3, 1999}
 
\authorrunning{S.Mereghetti et al. }
\titlerunning{ \rxj}
\maketitle
 
\begin{abstract}
We report \RXTE\ and \BSAX\ observations of the X--ray pulsar  \rxj, 
the neutron star/Be binary system with the longest known spin period.

Data obtained one month after the most recent outburst (July 1997),
show that the source has returned to its normal luminosity of a few
10$^{34}$ erg s$^{-1}$ while spin down continued at an average rate
of $\sim 5\times 10^{-8}$ s s$^{-1}$.

 
\end{abstract}

\section{Introduction}

\rxj is an accreting neutron star with a $\sim$25 min spin period, the longest known
period of any X-ray pulsar in a Be-star system.  This fact was realized
(Mereghetti, Stella \& De Nile 1993) only after the re--discovery of this source
in the ROSAT All Sky Survey and its identification with the 11$^{th}$ magnitude
Be star LS~I~+61$^{\rm o}$~235 (Motch et al.  1991).  Indeed the 25 min
periodicity had already been discovered with EXOSAT (White et al.  1987), but it
was attributed to a nearby source (4U 0142+614; see also Israel, Mereghetti \&
Stella 1994, Hellier 1994).

The optical companion of \rxj was classified as
B5IIIe (Slettebak 1985), but Coe et al. (1993) derived a somewhat earlier
spectral type of  O9--B0. This star is probably a member of the open
cluster NGC 663  at a distance of about 2.5~kpc (Tapia et al.  1991). For
this distance, the 1--20~keV luminosity during the EXOSAT detection in
1984 was $\sim 10^{36}$~erg~s$^{-1}$ (Mereghetti, Stella \& De Nile
1993).
 
All the observations of \rxj carried out after its re-discovery yielded
lower luminosities, of the order of a few $10^{34}$~erg~s$^{-1}$
(Hellier 1994, Haberl et al. 1998), until an observation with the
\RXTE\ satellite showed that in July 1997 the flux started to rise again
(Haberl, Angelini \& Motch 1998),
though not up to the level of the first EXOSAT observation. 
 
Here we report on observations performed with the \RXTE\ (Bradt et al.  1993)
and \BSAX\ (Boella et al.  1997a) satellites from 1996 to 1998.
 
\section{\RXTE\ Observations  }
  
\subsection{Spectral Analysis}
 
\rxj was observed with the \RXTE\ satellite on March 28, 1996 
from 11:17 UT to 22:11 UT. 
The results presented here are based on data 
collected with the Proportional Counter Array (PCA, Jahoda et al. 
1996). The PCA instrument consists of an array of 5 
proportional counters operating in the 2--60 keV energy range, 
with a total effective area of approximately 7000~cm$^{2}$  and a 
field of view, defined by passive collimators, of $\sim1\deg$   
FWHM.

For the analysis of the PCA data, time intervals with a source elevation
angle greater than 10$\deg$ were selected, resulting in a net exposure time of
about 18.5 ks.  Only the top layer anodes of the five proportional counters 
were
used to accumulate spectra and light curves of \rxj .  The background was
estimated using the latest version of the program PCABACKEST applicable to
the time period of our observation  
(version 1.5).  To account for the uncertainties in the response matrix (version
2.2.1), we added a 2\% systematic error to the PHA data.
 
Attempts to fit the PCA spectrum with single component models gave unacceptable
results, due to the presence of some contamination from the nearby source
4U~0142+614.  In fact the PCA collimator (Strohmayer \& Jahoda 1998) has a
transmission of about 58\% at the off-axis angle of this pulsar ($\sim$24').
Since the luminosity of 4U~0142+614 shows little or no variability on long
timescales (Israel \etal 1999a) we can estimate its contribution and model it in
the spectral analysis.  Assuming the spectral parameters measured with ASCA
(White et al.  1996; N$_H= 0.95\times 10^{22}$~cm$^{-2}$, power law photon index
$\alpha$ = 3.67, blackbody temperature kT = 0.386 keV) and reducing the
normalization according to the collimator response, we expect from 4U~0142+614
$\sim$10 counts s$^{-1}$ in the 2-20 keV range.  This corresponds to about 18\%
of the background-subtracted counts detected by the PCA in the same energy
range.  Due to the soft spectrum of 4U 0142+614, its contribution rises to over
35\% of the counts in the 2-4 keV range.  Therefore we included in the fits the
contribution from 4U 0142+614 with spectral parameters fixed at the above
values.

Even taking the contribution from 4U 0142+614 into account, a single power law
cannot describe the spectrum of \rxj.  The residuals show clear evidence for an
iron K emission line and for a spectral turnover at energies above $\sim$15 keV.
A better fit can be obtained by adding to the model a Gaussian emission line and
an exponential cut-off (or a broken power law) to the model.  The best fit
spectrum is shown in Fig.~1 and its parameters are reported in Table 1.

\begin{figure}[htb] 
\mbox{} 
\vspace{7.5cm} 
\includegraphics{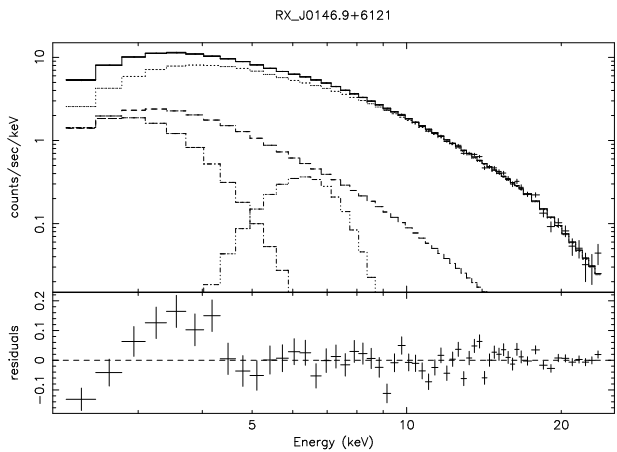} 
 \caption[]{ The \RXTE\ PCA best fit spectrum and its different model components,
corresponding to the parameters in Table 1.  Residuals are shown in the lower
panel.  }
 \label{nttvla} 
\end{figure}
 
\begin{table}
\begin{center}
\caption[]{\label{mag} Results of the \RXTE\ PCA spectral analysis.  All errors
represent 90\% confidence limits after introducing the 2\% systematic error.}
\begin{tabular}{ccc}
\hline
         &                 &                     \\
\hline
  \rxj   & photon index            & 2.05$^{+0.04}_{-0.43}$          \\
         & N$_H$ (atoms cm$^{-2}$) & $<$2.2 10$^{22}$                \\
         &   E$_{cutoff}$  (keV)   & 17.2$\pm$0.9                    \\
         &   E$_{fold}$  (keV)     & 7.0$^{+4.1}_{-2.5}$             \\
         &                         &                                 \\
   line  &   E   (keV)             &  6.3$^{+0.4}_{-1.7}$            \\
         & $\sigma$ (keV)          &  0.8$^{+2.5}_{-0.5}$            \\
         &  EW (eV)       & 168      \\
         &                 &                                        \\
4U~0142+614 & N$_H$ (atoms cm$^{-2}$)  &  0.95 10$^{22}$       \\
 {\it (fixed}    & photon index    &  3.67             \\
{\it parameters) }   & power law norm  &  0.09             \\
             & blackbody kT    &  0.386 keV        \\
             & blackbody norm  &  334 km$^2/10$ kpc$^{2}$    \\
             &             &                       \\
 \rchisq/dof &             &    1.24/50                \\
         
\hline
\end{tabular}
\end{center}
\end{table}

Since \rxj lies at low galactic latitude (l=129$^{\rm o}$.5, b=--0$^{\rm o}$.8)
it is likely that the observed iron line result from the diffuse emission from
the galactic ridge (Koyama et al.  1989; Yamauchi \& Koyama 1993).  Based on 
the
work by Yamauchi \& Koyama we derived the flux from the diffuse line
emission expected within the field of view of the PCA instrument.  Though
several uncertainties are involved in this estimate and the parameters of the 
Fe
line reported in Table 1 are affected by the uncertainty in the PCA calibration
around the Xenon L edge (at $\sim$5.5 keV; see, e.g., figure 1 in Dove et al
1998), we find that the observed Fe-line flux can be easily accounted for.  
This
conclusion is further supported by the absence of line emission in the \rxj
spectrum obtained with the ASCA imaging instruments (Haberl, Angelini \& Motch
1998), which allow a better subtraction of the local background.

The flux of \rxj in the 2-20 keV range corresponding to the best fit parameters
is $1.3\times10^{-10}$ erg cm$^{-2}$ s$^{-1}$ ($1.4\times10^{-10}$ erg
cm$^{-2}$ s$^{-1}$, corrected for the absorption).  For a distance of 2.5 kpc 
this
corresponds to a luminosity of $\sim$10$^{35}$ erg s$^{-1}$.

\subsection{Timing Analysis}
 
To derive the pulse period of \rxj we used a standard folding technique,
obtaining a value of 1407.8 $\pm$ 1.3 s.  The light curve, shown for 
three
different energy ranges  in Fig.~2,  has a    single broad peak,
as seen in previous observations (ASCA, ROSAT, see Haberl, Angelini \&
Motch 1998; \RXTE\, see Haberl et al.  1998). 
Moreover, thanks to the considerably higher signal to noise ratio,
some  substructures are clearly visible in our data,
as well as significant  pulse to pulse variability.
This is shown in Fig.~3 where we have plotted most of the individual pulses
visible in our observation. 

\begin{figure}[htb]
\mbox{}
\vspace{7cm}
\includegraphics{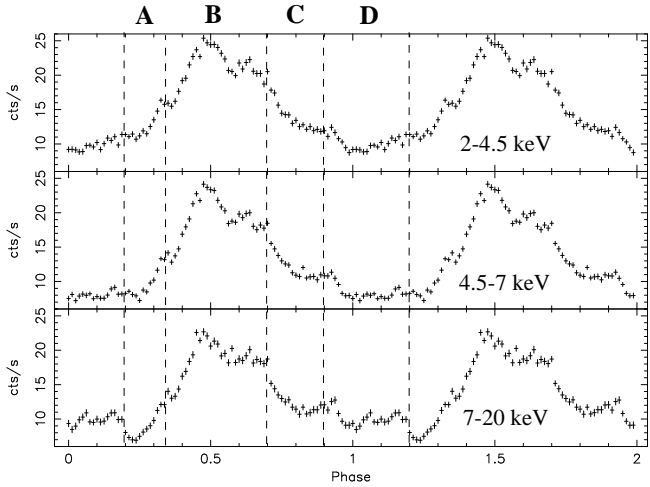}
\caption[]{ The \RXTE\ PCA light curve of \rxj folded at its 1407.8 s period, 
plotted in 3
different ranges of energy. The contribution of the background and of 
4U~0142+614
has been subtracted.  The vertical lines show the 4 phase
intervals chosen for the pulse phase spectroscopy.  }
\label{nttvla} 
\end{figure}
 
\begin{figure}[htb] 
\mbox{} 
\vspace{10cm} 
\includegraphics{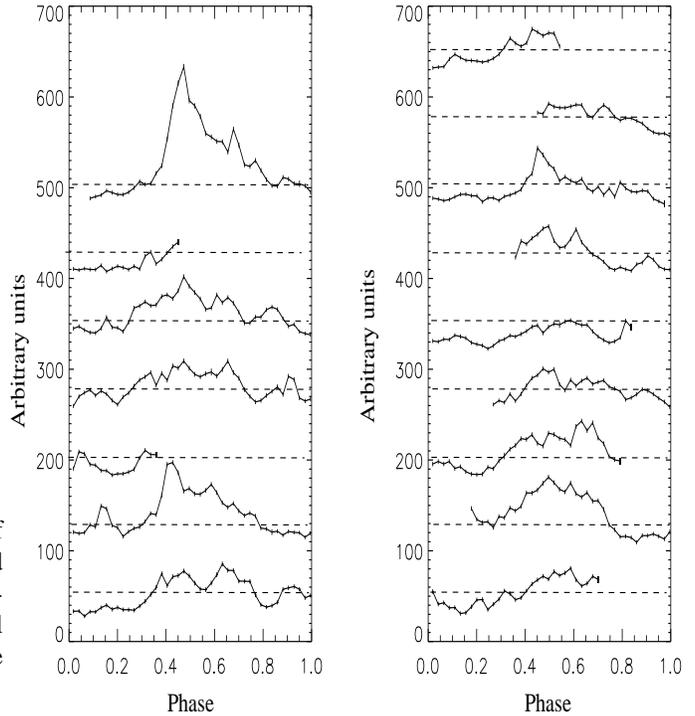} 
\caption[]{Light curves of individual pulses of \rxj. 
The time sequence  of the pulses is from bottom-left  to top-right.
Each pulse has been offset by an arbitrary constant, but the ordinate 
scale is the same for all the curves. The dashed lines correspond to the
average count rate of the whole observation (55 counts s$^{-1}$).}
 \label{nttvla} 
\end{figure}

\subsection{Phase-resolved Spectroscopy}
 
Some evidence for spectral variations as a function of the spin period phase are
apparent from Fig.~2.  To investigate this in more detail, we extracted
background subtracted spectra corresponding to the four phase intervals marked
in Fig.2.  As a first step, we fitted the total spectrum with the model of Table
1 and then, keeping all other parameters fixed, we renormalized the power
law representing \rxj for each of the four phase intervals; the ratios of the
observed spectra to these renormalized average models are shown in Fig.~4.  Some
features are clearly seen in these plots:  (1) a considerable excess above 10
keV during the interval D, (2) a softening of the whole spectrum during the
interval A, (3) a slight variation of the intermediate energy structure with
phase.

To investigate which components of the spectrum are phase dependent, the
four spectra were fitted with the   model discussed above, keeping the
parameters corresponding to the 4U~0142+614 contamination and to the Fe-line
fixed.  The results are reported in Table 2.  
Significant variations are seen in the power law spectral index
and absorption, while the errors in the cut-off parameters
are too large to establish a definite variation in the
different pulse phase intervals.
Note that during phase interval D, we obtain 
an upper limit to the 
photoelectric absorption  lower than the value expected
for \rxj from optical observations (Motch et al.  1997).
This is an indication that the   model
used is oversimplified and more spectral components
are probably required, at least at certain phase intervals.

\begin{figure}[htb] 
\mbox{} 
\vspace{7.5cm} 
\includegraphics{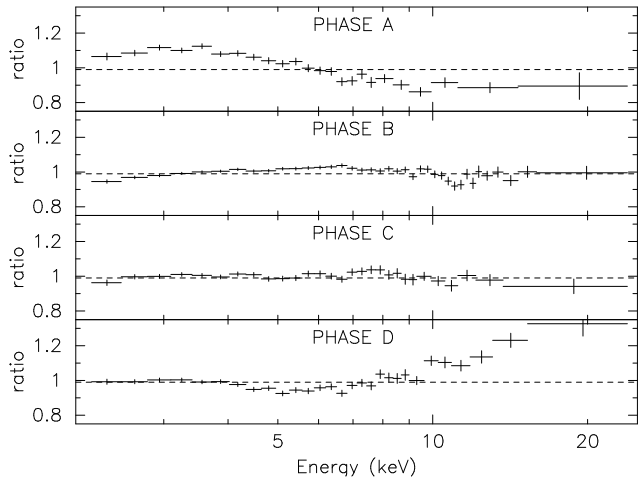} 
\caption[]{ Ratios of the   spectra corresponding to the    phase intervals 
shown in
Fig.2 to the average model spectrum.  The model is a power law with low-energy
absorption and high-energy cutoff (and the fixed components discussed 
in
the text to account for  4U~0142+614 and the galactic ridge iron line). 
The parameters are
fixed at the best-fit values for the total spectrum, except for the
normalization that is the best-fit value for each phase interval.  }
\label{nttvla} 
\end{figure}

\begin{table}
\caption[]{\label{mag} Results of the \RXTE\ PCA phase-resolved spectroscopy.  
All errors
represent 90\% confidence limits after intoducing the 2\% systematic error.}
\begin{center}
\begin{tabular}{cccc}
\hline
         &                 &                &       \\
\hline
 Phase A &   phot.index    & 2.27$\pm$0.06          &       \\
         &   N$_H$         & 1.4$\pm$0.5 10$^{22}$     &       \\
         &   E$_{cutoff}$  & 17.6$^{+3.9}_{-3.2}$ keV       &       \\
         &   E$_{fold}$    & 4.6$^{+34.5}_{-4.6}$ keV       &       \\
         & \rchisq/dof     &    0.90/53          &       \\
         &                 &                &       \\
 Phase B &   phot.index    & 2.14$\pm$0.03          &       \\
         &   N$_H$         & 2.7$^{+0.4}_{-0.3}$ 10$^{22}$     &       \\
         &   E$_{cutoff}$  & 17.3$^{+1.3}_{-1.5}$ keV       &       \\
         &   E$_{fold}$    & 8.5$^{+8.8}_{-4.1}$ keV        &       \\
         & \rchisq/dof     &    0.81/53          &       \\
         &                 &                &       \\
 Phase C &   phot.index    & 2.08$\pm$0.05          &       \\
         &   N$_H$         & 2.1$\pm$0.4 10$^{22}$     &       \\
         &   E$_{cutoff}$  & 16.6$^{+1.9}_{-1.7}$ keV       &       \\
         &   E$_{fold}$    & 5.0$^{+7.1}_{-3.4}$ keV        &       \\
         & \rchisq/dof     &    0.85/53          &       \\
         &                 &                &       \\
 Phase D &   phot.index    & 1.77$^{+0.04}_{-0.03}$    &       \\
         &   N$_H$         & $<$ 0.4 10$^{22}$     &       \\
         &   E$_{cutoff}$  & 17.7$^{+2.3}_{-2.2}$ keV       &       \\
         &   E$_{fold}$    & 7.8$^{+19.6}_{-5.1}$ keV       &       \\
         & \rchisq/dof     &    0.93/53          &       \\
         &                 &                &       \\
 
\hline
\end{tabular}
\end{center}
\end{table}

\section{\BSAX\ Observations}
 
We have analyzed the data
from the the Low--Energy Concentrator Spectrometer (LECS; 0.1--10~keV; 
Parmar et al. 1997) and Medium--Energy Concentrator Spectrometer 
(MECS; 1.3--10~keV; Boella et al. 1997b).
The MECS instrument consists of three identical grazing incidence telescopes
with imaging gas scintillation proportional counters in their focal planes.
The LECS instrument uses an identical concentrator system as the MECS, but 
utilizes an
ultra--thin (1.25~$\mu$m) detector entrance window and a driftless 
configuration to extend the low--energy response to 0.1~keV. The fields of view
of the LECS and MECS are circular with diameters of 37\arcmin\ and 
56\arcmin\, respectively. The energy resolution of both instruments is
$\sim$8.5$\sqrt{6/{\rm E_{keV}}}$\% FWHM.
In the overlapping energy range, the 
angular resolution of both instruments is similar and corresponds to 
90\% encircled energy within a radius of 2\farcm5 at 1.5~keV.
 
The \rxj position is included at various offset angles in four \BSAX\
observations of 4U\,0142+614 (see Table\,3).  The
original project was to monitor \rxj in order to remove the high energy
contribution of this pulsar from that of 4U\,0142+614 in the non--imaging
instruments.  The spectral and timing results we obtained for 4U\,0142+614 are
reported elsewhere (Israel \etal 1999a).

After the first observation, one of the three MECS units failed (1997 May 9) and
the data of the following pointings were obtained through the remaining two
MECS units.

\subsection{Spectral Analysis}
 
Spectra were extracted from circular regions with radius of 4\arcmin\ for both
the LECS and MECS instruments and rebinned so as to have $>$30 counts in each
energy bin to allow the use of $\chi^2$ statistics.  Spectral response matrices
appropriate to the particular source off--axis positions were used.  These
matrices include also a correction for the presence of the strongback used to
support the detector entrance window.  The background subtraction was performed
using source--free regions of each observation at the appropriate off--axis and
azimuth angle (to properly take into account the effect of the strongback).  We
used only MECS counts in the energy range 1.8--10~keV.  The background
subtracted MECS count rates for each observations are reported in Table\,3.  All
energy bins the count rate of which was consistent with zero were
not used in the spectral analysis.

\begin{table*}
\caption[]{\BSAX\ Observation log}
\begin{center}
\begin{tabular}{ccccc}
\hline 
Start Time & Stop Time & MECS Exposure   & Count rate & Off--axis   \\
 (UT)      & (UT)      & Time (s)        & MECS cts/s & angle (\arcmin\ ) \\ 
\hline
03--Jan--97~~05:47:20 & 04--Jan~~02:08:50 & 48226 & 0.156$\pm$0.002$^a$ &20 \\
09--Aug--97~~23:12:45 & 10--Aug~~07:37:40 & 16757 & 0.132$\pm$0.003     &20 \\
26--Jan--98~~12:13:02 & 27--Jan~~00:22:43 & 21785 & 0.214$\pm$0.004     &3  \\
03--Feb--98~~09:11:10 & 04--Feb~~02:36:39 & 31150 & 0.154$\pm$0.003     &20 \\ 
\hline
\hline
\end{tabular}\\
$^a$ Three MECSs on and the source is under the strongback.
\end{center}
\end{table*}

The January 1998 observation, during which \rxj was observed nearly on-axis,
provided the data of best quality.  In the limited energy range covered by the
LECS and MECS instruments, a power law model provides a satisfactory fit to the
spectrum of \rxj.  The best fit parameters are photon index of $\alpha = 1.67
\pm 0.1$, N$_H$ = (1.2$\pm$0.3) 10$^{22}$ cm$^{-2}$, corresponding to a flux of
3.1 10$^{-11}$ erg cm$^{-2}$ s$^{-1}$ (2-10 keV, corrected for the absorption).
The other three \BSAX\ observations gave consistent results in terms of
intensity and spectral parameters, to within the large uncertainties related to
the calibrations at off-axis angles.

\subsection{Timing Analysis}
The arrival times of the 1--10~keV photons from \rxj 
were corrected to the barycenter of the solar system and  
then used to determine the pulse period. 
A phase fitting technique was applied in order to better determine the pulse 
period during each observation; the results are reported in Table 4 
(1 $\sigma$ uncertainties are also given). 
The  light curves  in different energy ranges, folded at the best periods, 
show a broad single--peaked profile with a pulsed 
fraction of $\sim$47\% which, to within the statistical uncertainties, is energy 
independent over the 1--10~keV range.

\begin{table*}
\caption[]{\label{ew} Period Measurements of \rxj  }
\begin{center}
\begin{tabular}{ccccc}
\hline
Observation      &  Period & Error  & SATELLITE    & Reference        \\
Date             &   (s)   &  (s)   &              &                    \\
 \hline
1984 August 27-28               &  1455     &   3    &  EXOSAT     & White et 
al. 1987 \\
1993 February 12-13             &  1412     &   4    &  ROSAT PSPC & Hellier 
1994      \\
1994 September 18-19            &  1407.4   &   3.0  &  ASCA       & Haberl et 
al. 1998 \\
1996 January 21 - February 28   &  1407.28  &   0.02 &  ROSAT HRI  & Haberl et 
al. 1998 \\
1996 March 28                   &  1407.8  &   1.3     &  \RXTE\        & This work  
 \\
1997 January 3                  &  1405.4   &   0.6  &  \BSAX\        & This work  
   \\
1997 July 4-10                  &  1404.2   &   1.2  &  \RXTE\        & Haberl, 
Angelini \& Motch 1998 \\
1997 August 9                   &  1403.9   &   1.4  &  \BSAX\        & This work  
      \\
1998 January 26                 &  1405.6   &   1.0  &  \BSAX\        & This work  
     \\
1998 February 3                 &  1401.6   &   1.4  &  \BSAX\        & This work  
     \\
   \hline
\end{tabular}
\end{center}
\end{table*}
 
\section{Discussion}
 
 All the spin period measurements of \rxj have been collected 
in Table 4 and are shown in Fig.~5.
The first two values, obtained with EXOSAT in 1984
and with ROSAT in 1993, imply a spin-up of at least $1.6\times10^{-7}$
s s$^{-1}$ (equivalent to a spin-up timescale of the
order of $\sim$300 years).
Note that this is a lower limit, since we do not know
at which time after the 1984 outburst the spin period
reached the smaller value measured in recent observations.
In contrast, all period measurements obtained after
the rediscovery of this source are consistent with 
small fluctuations around  a smaller average spin-up.
A linear fit to the 1993-1998 data yields a value 
of  $5.4\times10^{-8}$s s$^{-1}$. 
This behaviour can be interpreted by assuming that the rapid spin-up
following the 1984 outburst was due to large angular 
momentum transfer through an accretion disk
in a phase of high accretion rate.
Indeed in 1984 \rxj had a luminosity of $\sim10^{36}$ erg s$^{-1}$,
while in all subsequent measurements it showed luminosities at least 
a factor ten smaller. This is illustrated in Fig.6, where
all published flux values are plotted as luminosities
for a distance of 2.5 kpc and after the appropriate conversions
to the 2-10 keV energy range.
 
A   flux increase of a factor 5
within one week was detected during the \RXTE\ observation of July 1997 
(Haberl, Angelini \& Motch  1998), indicating the possible start
of a new outburst from \rxj . 
As a consequence one might have expected a corresponding increase in
the spin-up. However, our  two \BSAX\ measurements,  obtained
respectively one and six months later, did not show any significant
variation in the spin period. This might be due to the fact that the
July 1997 outburst was smaller (in peak luminosity and possibly 
in length, see Fig. 6) than the 1984 one. 
 
Though our folded light curve of \rxj is similar to those reported from previous
observations, the better statistics of our data allows to confirm the presence
of significant substructure in the pulse profile.  Examples of this are the small
dip in the main broad peak occurring at phase $\sim$0.6 and the structure in 
phase interval D, which is visible mostly in the higher energy range.  Note that,
although \rxj was brighter during the July 1997 \RXTE\ observation reported by
Haberl et al.  (1998), on that occasion the source was observed at a large
offset angle resulting in a reduced count rate.

The good statistics also permitted, for the first time in this source,
to detect variations of the energy spectrum as a function of the pulse
period phase. 
  
We finally note that, contrary to previous reports   we do
not find any evidence for a spectral cut-off at low energy. 
A spectral cut-off at $\sim$4 keV  was obtained by a combined spectral fit of
non simultaneous \RXTE\ and ASCA data (Haberl, Angelini \& White 1998)
and interpreted as possible evidence for a relatively weak magnetic field
of the order of a few 10$^{11}$ G. Our results indicate a cut-off value
in the range usually observed in X--ray pulsars.

Until recently \rxj and X Persei had extreme properties among Be X-ray pulsars,
by virtue of their long spin period and relatively low luminosity.  New X--ray
and optical observations led to the discovery of a few other systems with
similar characteristics:  RX J0440.9+4431, RX J1037.5--564 (Reig \& Roche 1999), 1WGA
J1958.2+3232 (Israel et al.  1998), and possibly 1SAX J0103.2--7209 (Israel et
al.  1999b).

These long period X-ray pulsating sources are characterized by relatively low and steady
luminosities, in contrast with the brighter systems the discovery of which has
been favoured in the past during their strong outbutsts.  This is illustrated in
Fig.~7, which shows the luminosity versus pulse period of the 44 X--ray pulsars
with Be companions currently known.  For \rxj and the other similar sources
mentioned above, we have indicated the observed range of luminosity values.  For
the other systems we have reported the maximum observed luminosity, since in
most cases only upper limits or very uncertain values exist for their quiescent
fluxes.  Note that fast spinning, magnetized neutron stars are not visible at
low luminosities, owing to the action of the centrifugal barrier in preventing
accretion onto the neutron star surface (Stella, White \& Rosner 1986).  On the
other hand, neutron stars with spin periods longer than a few hundreds seconds
can easily display pulsed emission at low accretion rates.  Future, more
sensitive observations will further increase this sample that probably
forms the bulk of the Be/neutron star binaries.

\begin{figure}[htb] 
\mbox{} 
\vspace{7.5cm} 
\includegraphics{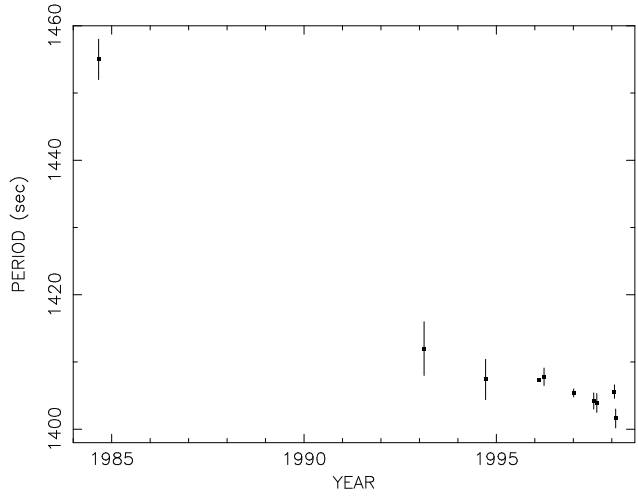} 
\caption[]{ Pulse period history of \rxj.  See Table 4 for details.  }
\label{nttvla} 
\end{figure}
  
\begin{figure}[htb] 
\mbox{} 
\vspace{7.5cm} 
\includegraphics{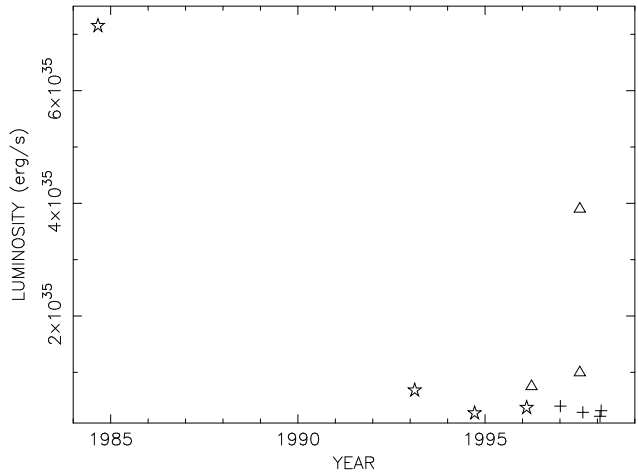} 

\caption[]{ Luminosity history of \rxj.  The luminosities are in the 2--10 keV
band and for a distance of 2.5 kpc.  Triangles and crosses refer to \RXTE\ and
\BSAX\ observations respectively while stars to data collected by other
satellites (EXOSAT, ROSAT and ASCA).  Except for the 1984 and July 1997
outbursts, all the flux measurements are consistent with an average luminosity
of $\sim 5\times 10^{34}$ erg~s$^{-1}$.  }

\label{nttvla} 
\end{figure}

\begin{figure}[htb] 
\mbox{} 
\vspace{7.5cm} 
\includegraphics{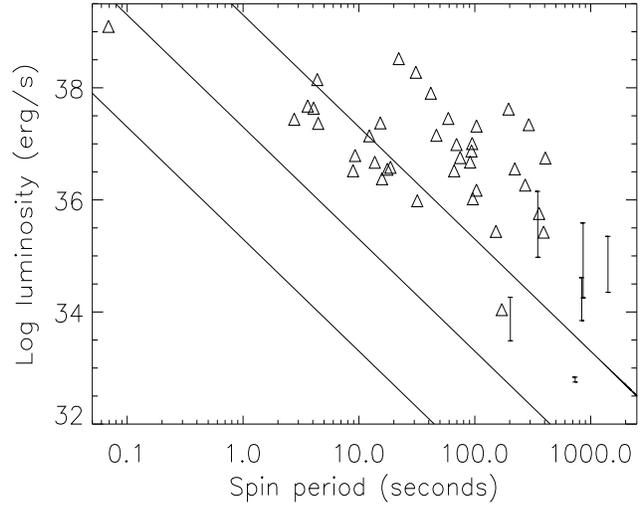} 

\caption[]{ Correlation between X-ray luminosity (in the range 2-10 keV) and
spin period of all the Be X-ray pulsar systems.  For transient sources
(triangles) only maximum luminosities are plotted.  The three lines correspond
to the centrifugal equilibrium (White, Nagase \& Parmar 1995) for different
values of the magnetic field (from left to right:  10$^{11}$, 10$^{12}$,
10$^{13}$ G).  }

\end{figure}


\end{document}